\documentclass[useAMS,onecolumn]{mn2e}
\usepackage{times}
\usepackage{rotating}
\long\def\symbolfootnote[#1]#2{\begingroup%
\def\thefootnote{\fnsymbol{footnote}}\footnote[#1]{#2}\endgroup}
\newcommand{\gae}{\lower 2pt \hbox{$\, \buildrel {\scriptstyle >}\over {\scriptstyle
\sim}\,$}}
\newcommand{\lae}{\lower 2pt \hbox{$\, \buildrel {\scriptstyle <}\over {\scriptstyle
\sim}\,$}}

\begin{document}

\title[Mild deviation from power-law distribution]{Evidence for mild 
deviation from power-law distribution of electrons in relativistic shocks: 
GRB 090902B}

\author[Barniol Duran \& Kumar]{R. Barniol Duran$^{1,2}$\thanks
{E-mail: rbarniol@physics.utexas.edu, pk@astro.as.utexas.edu}
and P. Kumar$^{2}$\footnotemark[1] \\
$^{1}$Department of Physics, University of Texas at Austin, Austin, TX 78712, USA\\
$^{2}$Department of Astronomy, University of Texas at Austin, Austin,
TX 78712, USA}

\date{Accepted; Received; in original form 2011 March 25}

\pubyear{2011}

\maketitle

\begin{abstract}
Many previous studies have determined that the long lasting emission
at X-ray, optical and radio wavelengths from gamma-ray bursts (GRBs), 
called the afterglow, is likely produced by the external forward shock model.  
In this model, the GRB jet interacts with the circum-stellar medium and 
drives a shock that heats the medium, which radiates via synchrotron emission.  
In this work, we carried out a detailed analysis of the late time afterglow data of
GRB 090902B using a very careful accounting of the Inverse Compton losses.  
We find that in the context of the external forward shock model, 
the only viable option to explain the X-ray and optical data of GRB 090920B 
is to have the electron energy distribution deviate from a power-law shape 
and exhibit some slight curvature immediately downstream of the shock front (we explored 
other models that rely on a single power-law assumption, but they all fail to explain the observations).
We find the fraction of the energy of shocked plasma in magnetic field to be $\sim 10^{-6}$ 
using late time afterglow data, which is consistent with the value obtained using 
early gamma-ray data. Studies like the present one might be able to provide a link between GRB
afterglow modeling and numerical simulations of particle acceleration in
collisionless shocks. We also provide detailed calculations for the early ($\lae 10^3$ s)
high energy ($> 100$ MeV) emission and confirm that it is consistent with origin in the 
external forward shock.  We investigated the possibility that the $\sim 10$ keV excess 
observed in the spectrum during the prompt phase also has its origin
in the external shock and found the answer to be negative.
\end{abstract}


\section{Introduction}

The external forward shock model (see, e.g., Rees \& M\'esz\'aros 1992; 
M\'esz\'aros \& Rees 1993, 1997; Paczy\'nski \& Rhoads 1993; 
Wijers, Rees \& M\'esz\'aros 1997; Sari, Piran \& Narayan 1998; 
Dermer \& Mitman 1999) has proven to be a very useful concept in the study of GRBs
(for a review of this phenomenon see, e.g., Piran 2004).  The
relativistic GRB jet or outflow interacts with the surrounding
medium of the progenitor star (circumstellar medium or CSM) and drives
a forward shock that accelerates the particles in the CSM, which 
radiate via synchrotron and Inverse Compton mechanisms.  By modeling
GRB afterglows one can
learn more about the CSM medium properties (its density),
general properties of the outflow (total kinetic energy in the 
shocked medium)
and some details of the shock process (fractions of total 
energy in the shocked fluid imparted to electrons and magnetic fields).
Particles are likely accelerated in collisionless shocks by the Fermi process
and the resultant electron energy distribution is expected to be a
decaying power-law in electron energy with index $p$ 
(see, e.g., Krymskii 1977, Axford, Leer \& Skadron 1978; 
Bell 1978; Blandford \& Ostriker 1978; 
Blandford \& Eichler 1987; Gallant \& Achterberg 1999; Achterberg et al. 2001;
Sironi \& Spitkovsky 2011).  Various studies have attempted
to calculate the value of $p$ and to test for its universality among 
all bursts (see, e.g.,
Shen, Kumar \& Robinson 2006).  Even though in the majority of previous studies 
a single value of $p$ is assumed for a particular burst, there is the possibility that the 
distribution function might deviate from a single power-law (see, e.g., Li
\& Chevalier 2001).

In this paper we study the late ($\gae$ 0.5 d) afterglow of GRB 090902B
in the context of the external forward shock model. In Section 2, we study the
afterglow data for this GRB.  In Section 3, we present 
alternatives to the scenario proposed in Section 2. In particular, 
we devote most of this section on a very detailed calculation of Inverse 
Compton losses, which might mitigate some of the issues presented in 
Section 2.  In Section 4, we show 
that the only viable solution seems to be that the electron energy
distribution is a little steeper for the X-ray radiating electrons than for the
optical electrons in the external forward shock, that is, there is some curvature in the
electron energy distribution spectrum.  In Section 5, we use the results 
of the previous section to study the early ($\sim 50$ s) high-energy
gamma-ray data of GRB 090902B and find it consistent with the same 
origin as the late time afterglow emission. In Section 6, we explore 
the possibility that the additional power-law spectral component 
found at $\sim 10$ s, in addition to the prompt Band function,
has also an external forward shock origin. We discuss 
our results in Section 7 and present our Conclusions 
in Section 8.

\section{Late time afterglow data of GRB 090902B}

GRB 090902B (Abdo et al. 2009) has been studied extensively for its 
high energy emission during the prompt phase 
detected by the LAT (Large Area Telescope) onboard of the {\it Fermi}
satellite (see, e.g., 
Asano, Inoue \& M\'esz\'aros 2010, Feng \& Dai 2010,
Ghisellini, Ghirlanda \& Nava 2010, 
Kumar \& Barniol Duran 2010, 
Toma, Wu \& M\'esz\'aros 2010, 
Liu \& Wang 2011, Zhang et al. 2011, Zhao, Li \& Bai 2011, 
Zou, Fan \& Piran 2011).  
In this paper we focus on the late time afterglow behavior in the 
radio, optical and X-ray bands (McBreen et al. 2010, Pandey et al. 2010, 
Cenko et al. 2011) with 
the assumption that radiation in these bands is produced by the synchrotron 
process in the external forward shock.  We study the afterglow data 
after about a day of the explosion, because the optical data previous 
to this epoch seems to be dominated by the external reverse shock (Pandey 
et al. 2010).  Here and in the rest of the paper, we use the convention that the 
observed specific flux at a particular energy, $\nu$, is given by 
$f_{\nu}(t) \propto \nu^{-\beta} t^{-\alpha}$, where $t$ is the observed
time and $\beta$ ($\alpha$) is the spectral (temporal) index.
The GRB 090902B observations for $t \gae 1$ d can be summarized as follows.
The optical data (2 eV), detected by UVOT (Ultraviolet Optical Telescope) 
onboard of the {\it Swift} satellite, 
shows $\beta_{opt}=0.82\pm0.10$ and $\alpha_{opt}=0.89\pm0.03$, 
whereas the X-ray data (1 keV), detected by XRT (X-ray Telescope)
also onboard {\it Swift}, shows $\beta_x = 0.90\pm0.13$ and $\alpha_x = 1.36\pm0.03$
(Cenko et al. 2011).  

We assume that electrons in the CSM are accelerated to a power-law
in the external forward shock model, such that the electrons energy distribution is 
given by $n(\epsilon)\propto \epsilon^{-p}$, where $p$ is the power-law index. 
In this scenario both the spectral index and the temporal decay index depend on 
$p$ (Sari et al. 1998), therefore, one can relate $\alpha$ and $\beta$ 
via the ``closure relations'' (see table 1 of Zhang \& M\'esz\'aros 2004).
The values of the spectral and decay indices will depend also on the region 
where the observed frequency falls in the synchrotron spectrum.  The synchrotron 
spectrum is characterized by three frequencies: the self-absorption frequency ($\nu_a$), 
the injection frequency ($\nu_i$), and the cooling frequency ($\nu_c$);  
$\nu_i$ and $\nu_c$ correspond to the synchrotron frequencies of electrons 
(just downstream of the shock front) whose energy correspond to the minimum energy of injected electrons 
and to the electrons that cool on a dynamical time, respectively, and $\nu_a$ is the highest 
frequency at which the system becomes opaque to synchrotron absorption.

Trying to explain both the X-ray and optical late time data for 
GRB 090902B in the context of the external forward shock model is not straightforward.
We use the common terminology ``slow cooling'' (``fast cooling''), for $\nu_i<\nu_c$
($\nu_c < \nu_i$), and constant density CSM (wind medium) for $s=0$ ($s=2$), where the CSM 
density falls off as $\propto R^{-s}$ and $R$ is the distance from the center 
of the explosion.  For this GRB we have the following options:

\noindent{\bf Option 1:} If the optical band, $\nu_{opt}$, and the X-ray band, $\nu_{x}$,
are in the same region of the synchrotron spectrum, that is, 
$\nu_i < \nu_{opt} < \nu_x < \nu_c$, then $\beta_x = \beta_{opt}$, 
which is supported by the data. However, the temporal decay indices in
these two bands, which should be exactly the same, are very different.
The X-ray decay is considerable steeper than the optical one, 
$\Delta \alpha = \alpha_x - \alpha_{opt} = 0.47 \pm 0.06$, and this 
particular issue is crucial for the rest of the paper.

\noindent{\bf Option 2:} If the optical band and the X-ray band lie in different parts 
of the spectrum, for instance, $\nu_i < \nu_{opt} < \nu_c < \nu_x$, 
then the spectral indices should differ by 
$\Delta \beta = \beta_x - \beta_{opt} = 0.5$, which is not supported 
by the data.   Moreover, $\Delta \alpha$ is expected to be 
$ -1/4$ (wind) or $1/4$ (constant 
density CSM), which is also inconsistent with the data -- the X-ray decays
too quickly.

\noindent{\bf Option 3:} If both optical and X-ray bands lie above the 
cooling frequency, that is, $\nu_i < \nu_c < \nu_{opt} < \nu_x$,  
the discrepancies with the expected data and the observations 
are similar to Option 1.  Since the temporal decay index is 
independent of type of medium there is no way to discriminate its 
type (Kumar 2000). This case is analogous to the fast cooling case where
$\nu_c < \nu_i < \nu_{opt} < \nu_x$.

\noindent{\bf Option 4:} Any possibility where $\nu_{opt}<\nu_i$
is ruled out by the data.  For the case of slow cooling, the spectrum 
would be $\beta_{opt}=-1/3$ for both types of medium, which is inconsistent 
with the observed optical spectrum. Moreover, the optical light curve should 
be slowly rising (flat) for the constant medium (wind) case, which is 
inconsistent with the decaying light curve.  For the case of fast cooling, 
if $\nu_{opt} < \nu_c < \nu_i$, then $\beta_{opt}=-1/3$, 
and if $\nu_c < \nu_{opt} < \nu_i$, then $\beta_{opt}=1/2$ and $\alpha_{opt}=1/4$ -
for both types of medium - which is inconsistent with the optical 
spectrum and light curve.  

We can see that Option 4 faces severe difficulties, thus we will not 
consider it any further. We will explore Options 1-3 in detail throughout the paper.   

First, we try to determine the type of medium that the blast wave
is running into.  For Options 1 and 2, the wind medium case can be ruled out.  
The reason is that, for Option 1, both the expected optical and X-ray decay would 
be too steep compared with the observations. For Option 2, one would expect 
$\Delta \alpha = \alpha_x - \alpha_{opt} = -1/4$, that is, the 
optical decay should be steeper than the X-ray one, which is the opposite 
to what it is seen.  For these reasons, the only 
viable possibilities left are Options 1 and 2 with constant CSM,
and for Option 3, optical and X-ray fluxes are independent of the type of CSM. 

Pandey et al. (2010) and Cenko et al. (2011) have also analyzed 
GRB 090902B and they both prefer Option 2 with a constant density CSM.  
Pandey et al. (2010) suggested a value of 
$p=1.8\pm0.2$, however, as the authors point out, the optical 
spectrum and X-ray temporal decay are inconsistent with the observed 
values.  They appeal to optical extinction, which would make the 
optical spectrum agree with the observed value; nevertheless, 
the expected X-ray temporal decay is shallower than the 
observed one.  Cenko et al. (2011) find $p=2.21\pm0.02$, which
naturally mitigates the problem Pandey et al. (2010) have with the
optical spectrum, but also gives an X-ray temporal decay that is shallower 
than the observed one.  They suggest radiative losses to make the X-ray 
decay steeper - consistent with observations - however, as we will 
see in the next section, this would also affect the optical decay, 
and therefore it is not a viable option.

\section{Saving the external forward shock model}

In this section, we explore a number of different possibilities that might modify the 
standard external forward shock model and help us reconcile the 
theory with the observations for GRB 090902B.  In particular, we are interested in 
mechanisms that could potentially make the X-ray light curve {\it steeper} 
than expected in the simple external forward shock model.  We consider 
the following possibilities: (i) Radiative losses in the blast wave, (ii) 
Temporal evolution of microphysical parameters, (iii) Temporal evolution 
of Compton-$Y$ parameter which would affect only the light curve of the 
observing band above $\nu_c$ and (iv) Curvature in the injected 
electron spectrum.   

These possibilities have been discussed in the literature
and applied to a number of GRBs.  Radiative losses in the blast wave was considered  
by, e.g., Cohen, Piran \& Sari (1998), Sari et al. (1998).  Also, 
the possibility of having the microphysical parameters vary with 
time in the external shock was proposed by Panaitescu et al. (2006).  Recently, Wang et al. (2010)
considered the possibility of having Klein-Nishina suppression weaken 
with time so as to increase the Inverse Compton losses and 
steepen the $> 100$ MeV light curve of {\it Fermi} GRBs.  
A curvature or steepening of the injected electron spectrum has been  
considered by, e.g., Li \& Chevalier (2001), Panaitescu (2001), Panaitescu \& Kumar (2001a), 
Grupe et al. (2010). We now explore possibilities (i)-(iv) to find out if any of 
these can help us understand the optical and X-ray afterglow data for GRB 090902B.

\subsection{Radiative losses or temporal evolution of microphysical parameters}

For a constant CSM, the specific flux for an observed band $\nu > \nu_i$ 
is given by (see, e.g., Sari et al. 1998, Kumar 2000, Panaitescu \& Kumar 2000)

\begin{equation} \label{eq:flux}
f_{\nu} \propto  \left\{ \begin{array}{ll}
E_{KE,iso}^{p+3 \over 4} \epsilon_e^{p-1} \epsilon_B^{p+1 \over 4} n^{1 \over 2} t^{-\frac{3(p-1)}{4}} \nu^{-\frac{p-1}{2}} & \textrm{if $\nu_i < \nu < \nu_c$} \\
E_{KE,iso}^{p+2 \over 4} \epsilon_e^{p-1} \epsilon_B^{p-2 \over 4} t^{-\frac{3p-2}{4}}  \nu^{-\frac{p}{2}} (1+Y)^{-1} & \textrm{if $\nu_c < \nu$}. \\
\end{array} \right.
\end{equation}
where $\epsilon_e$ and $\epsilon_B$ are the fractions of energy of the shocked gas in electrons
and magnetic fields, respectively, $t$ is the time since the beginning of the explosion in 
the observer frame,  $E_{KE,iso}$ is the isotropic kinetic energy in the shocked medium, $n$ 
is the density of the CSM and $Y$ is the Compton-$Y$ parameter, which is the ratio of the 
Inverse Compton to the synchrotron loss rates.  Equation (\ref{eq:flux})
is valid for $p>2$; closure relations for $p<2$ can be found in table 1 of Zhang \& M\'esz\'aros (2004).

According to Option 2, for $p=2.2$, the expected optical flux ($\nu_i < \nu_{opt} < \nu_c$)
is given by $\propto t^{-0.90}\nu^{-0.60}$, consistent with both the optical decay 
and spectrum within 1-$\sigma$ and $\sim2$-$\sigma$, respectively, 
while the expected X-ray flux ($\nu_c < \nu_x$) is $\propto t^{-1.15}\nu^{-1.10}$,
consistent with the observed X-ray spectrum within 2-$\sigma$, however,  
inconsistent with the observed X-ray decay by more than 5-$\sigma$. To be 
consistent, the  X-ray light curve must be steepened by $\propto t^{-0.21}$.

If the X-ray band is above $\nu_c$ then, according to the second part 
of equation (\ref{eq:flux}), the way to steepen the light curve is 
by appealing to a decrease with time of $E_{KE,iso}$, $\epsilon_e$, $\epsilon_B$ or 
an increase of $Y$.  The decrease of $E_{KE,iso}$, $\epsilon_e$ or $\epsilon_B$ will also steepen 
the optical light curve, therefore, the increase of Compton-$Y$ is the 
only possibility that we consider.  Let us explore these arguments in detail now. 

Radiative losses make the kinetic energy in the external 
forward shock decrease with time.  For the X-ray band ($\nu_c < \nu_x$) with $p=2.2$, the observed flux is 
$\propto E_{KE,iso}^{1.05}$ and, thus, $E_{KE,iso}$ should decrease as 
$\propto t^{-0.2}$ to steepen the X-ray value to the observed value.
However, since the flux in the optical band ($\nu_i < \nu_{opt} < \nu_c$) is $\propto E_{KE,iso}^{1.3}$, 
the optical light curve will steepen by $\propto t^{-0.26}$, making the 
optical decay $\alpha_{opt} = 1.16$ inconsistent with the observations by 
more than 5-$\sigma$. Therefore, radiative losses cannot save the external forward shock model in 
Option 2. 

Appealing to a temporal evolution of $\epsilon_e$ or $\epsilon_B$ faces 
similar difficulties.  The reason is that at $>0.5$ d, 
for electrons radiating at X-ray and optical bands, Compton-$Y$ is 
$Y\gae$ few, and it roughly behaves as $Y \propto \epsilon_e^{1/2} \epsilon_B^{-1/2}$; 
the dependence of flux on $\epsilon_e$ -- and also $\epsilon_B$ -- turns out to be similar 
below and above $\nu_c$.  Therefore, appealing to a temporal evolution of $\epsilon_e$ or 
$\epsilon_B$ cannot steepen the X-ray light curve to the observed value, 
and at the same time leave the optical light curve decay unchanged\footnote{
Conversely, also in the context of Option 2, one could fix {\it first} the 
X-ray band with $p=2.5$, but in this case the 
optical light curve would need to be made shallower by $\propto t^{0.24}$.
The same arguments as above prevent time-varying $E_{KE,iso}$ or 
microphysical parameters to achieve this.}.


In Options 1 and 3, the X-ray and optical bands lie in the same spectral regime, 
therefore their fluxes have the exact same dependence on energy and microphysics parameters.  Appealing 
to a temporal change of any of these would modify both light curves exactly 
the same way.  For this reason, neither radiative losses nor temporal evolution of microphysical parameters
can explain why the X-ray light curve decays faster than the optical one does.  

\subsection{Temporal evolution of Compton-$Y$}

As mentioned above, another possibility could be that the Compton-$Y$ parameter 
increases with time as $(1+Y)\propto t^{0.21}$.  Since $Y$ only affects the 
flux for $\nu_c < \nu$ -- see equation (\ref{eq:flux}) -- 
it would only affect the X-ray band, not the optical one (when $\nu_i < \nu_{opt} < \nu_c$), 
and since the optical data already agrees 
with the expected value, then this possibility is very attractive.

The Compton-$Y$ parameter that needs to be calculated is the one 
for electrons radiating at 1 keV, $Y_x$, since it is the 
X-ray flux at 1 keV that needs to be steepened.  Moreover, the temporal 
behavior $(1+Y_x)\propto t^{0.21}$ should hold during 
the entire period of X-ray observations, which start at $\sim 12.5$ h and extend 
until $\sim 15$ d.  
In the following subsections, we search the 4-D parameter space 
($\epsilon_e$, $\epsilon_B$, $n$, $E_{KE,iso}$) to determine if there is any part of the 
parameter space that satisfies the condition that $(1+Y_x) \propto t^{0.21}$.  
Before this, we present a very detailed calculation of Compton-$Y$ parameter
for electrons of arbitrary Lorentz factor (LF), where we take into account the 
effect of Inverse Compton and synchrotron losses on the electron energy 
distribution self-consistently and include the Klein-Nishina cross section
and electron recoil effects on Compton scatterings.

\subsubsection{Calculation of Inverse Compton loss}

A general calculation of Inverse Compton loss has been carried out by a number of authors
(see, e.g., Jones 1968, Blumenthal \& Gould 1970, see also, Blumenthal 1970).  Inverse Compton loss 
has also been calculated in the context of GRB prompt emission and afterglows
(see, e.g., Panaitescu \& Kumar 2000; Sari \& Esin 2001; and more recently  
Bo\v{s}njak, Daigne \& Dubus 2009, Nakar, Ando \& Sari 2010, Wang et al. 2010,
Daigne, Bo\v{s}njak \& Dubus 2011).  
The calculation of Inverse Compton loss, which allows us to calculate Compton-$Y$,
is not straightforward. Electrons cool via Inverse Compton scattering when they interact with synchrotron 
photons, and the same electron population is the one that emits the synchrotron 
photons.  It is a problem of feedback, because the electron population in turn 
depends on the cooling the electrons experience.  In this subsection we outline 
the calculation for Compton-$Y$.  We include Klein-Nishina effects and also 
relativistic corrections of the outgoing energy of the Inverse Compton scattered 
photons.

\begin{figure}
\begin{center}
\includegraphics[width=8cm, angle=0]{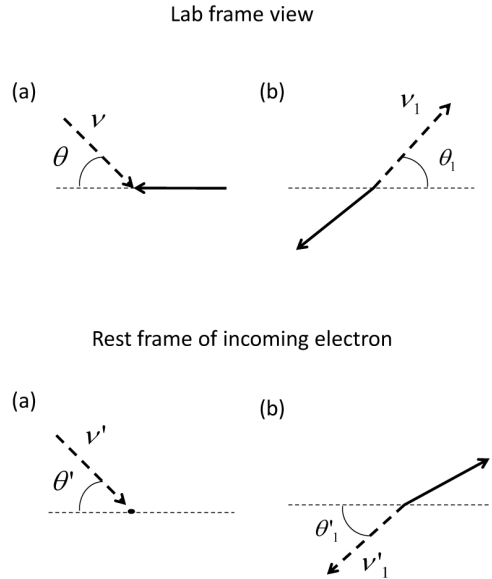}
\end{center}
\caption{ Geometry for scattering of a photon by an electron.  The 
electron trajectory is denoted by a thick solid line, while 
the trajectory of the incoming and outgoing photon is denoted 
by a thick dashed line.  We present the scattering viewed 
from the lab frame (top) and viewed from the rest frame of the incoming
electron, that is, the electron is at rest in this frame before scattering 
(bottom).  The diagrams before and after the collision are presented 
in (a) and (b), respectively.  The un-primed quantities are 
in the lab frame, while the primed ($'$) quantities are in the 
electron rest frame before scattering.}
\label{fig1} 
\end{figure}

In the co-moving frame of the electron, the energy of an outgoing photon 
after the scattering off of an electron is given by (see, e.g., Rybicki \& Lightman 1979, 
hereafter RL79)

\begin{equation} \label{eq:IC}
x'_1 = \frac{x'}{1+x'(1-\cos{(\theta'_1 - \theta')})},
\end{equation}
where the subscript ``1'' corresponds to the outgoing photon 
and the superscript ($'$) means that the quantity is measured in the 
co-moving frame of the electron before scattering.  The quantity $x'$ ($x'_1$) is defined 
as $x' = h \nu'/m_e c^2$ ($x'_1 = h \nu'_1/m_e c^2$), where $\nu'$ ($\nu'_1$)
is the frequency of the incoming (outgoing) photon, and $h$, $c$ and $m_e$ 
are Planck's constant, the speed of light and the mass of the electron, 
respectively.  The angle $\theta'$ ($\theta'_1$) is that of the direction 
of the incoming (outgoing) photon with respect to the direction of the 
electron's momentum before scattering (see Figure \ref{fig1}).  

The frequency of the incoming photon in the electron rest frame before 
scattering can be expressed in terms of the frequency measured in 
the lab frame, $\nu$, by using the relativistic Doppler formula 

\begin{equation} \label{eq:Dopler1}
\begin{array}{lcl}
\nu' & = & \nu \gamma_e (1 + \beta_e \cos \theta)    \\
\nu & = & \nu' \gamma_e (1 - \beta_e \cos \theta') 
\end{array}
\end{equation}
and the transformation between the angle of the incoming photon
$\theta'$ in the electron rest frame and the lab frame, $\theta$, 
is given by 

\begin{equation} \label{eq:Dopler2}
\sin \theta' = \left( \frac{\nu}{\nu'} \right) \sin \theta = \frac{\sin \theta}{\gamma_e (1 + \beta_e \cos \theta)}
\end{equation}
where $\beta_e$ is the velocity of the electron, $v_e$, divided by $c$, 
$\gamma_e = 1/\sqrt{1-\beta_e^2}$ is the LF of the electron.

The energy in a bundle of scattered photons in the electron rest frame is given by 

\begin{equation} \label{eq:dEdt1}
dE'_s = \left[ d\Omega' d\nu' \frac{I'_{\nu'}(\theta')}{h \nu'} \right] \frac{d\sigma_{KN}}{d\Omega'_1} h \nu'_1 d\Omega'_1 dt'
\end{equation}
where $I'_{\nu'}$ is the specific intensity (in units of 
erg s$^{-1}$ cm$^{-2}$ sr$^{-1}$ Hz$^{-1}$), $d\Omega'$ ($d\Omega'_1$)
is the differential solid angle of the incoming (outgoing) photons, 
$d\sigma_{KN}/d\Omega'_1$ is the differential cross section (the 
Klein-Nishina formula) and $dt'$ is the duration of the event 
measured in the electron rest frame. The quantity in the square 
bracket is the number of photons per unit time per unit area 
moving within solid angle $d\Omega'$ and frequency band $d\nu'$ 
incident on the electron.

The component of the momentum vector of the scattered photons along 
the electron velocity considered above is

\begin{equation}
dp'_s = d\Omega' d\nu'	\frac{I'_{\nu'}(\theta')}{h \nu'} \frac{d\sigma_{KN}}{d\Omega'_1} \frac{h \nu'_1}{c} \cos \theta'_1 d\Omega'_1 dt'.
\end{equation}

The scattered photon energy in the lab frame is given by (see, e.g., 
RL79)

\begin{equation}
dE_s = \gamma_e (dE'_s - v_e dp'_s) = \gamma_e (1 - \beta_e \cos \theta'_1) dE'_s.
\end{equation}
Using the fact that the time interval in the lab frame is $dt = \gamma_e dt'$, 
one finds

\begin{equation}
\frac{dE_s}{dt} = (1 - \beta_e \cos \theta'_1) \frac{dE'_s}{dt'},
\end{equation}
and combining this last equation with equation (\ref{eq:dEdt1}), we arrive 
to  

\begin{equation}
\frac{dE_s}{dt} = \int \! d\Omega' d\nu' \frac{I'_{\nu'}(\theta')}{\nu'} \int \! d\Omega'_1 \nu'_1 \frac{d\sigma_{KN}}{d\Omega'_1} (1 - \beta_e \cos \theta'_1).
\end{equation}
In this last equation, both $d\sigma_{KN}/d\Omega'_1$ and $\nu'_1$ are functions 
of $\nu'$ and $(\theta'_1 - \theta')$, and we can use equation (\ref{eq:IC})
to eliminate $\nu'_1$.  We thus find a general equation that 
describes the energy loss of an electron due to Inverse Compton cooling

\begin{equation} \label{eq:generaldEdt}
\frac{dE_s}{dt} = \int \! d\Omega' d\nu' I'_{\nu'}(\theta') \int \! d\Omega'_1 \frac{1 - \beta_e \cos \theta'_1}{1+x'[1 - \cos (\theta'_1 - \theta')]} \frac{d\sigma_{KN}}{d\Omega'_1},   
\end{equation}
where the Klein-Nishina cross section is given by (see, e.g., RL79
eq. 7.4) 

\begin{equation} 
\frac{d\sigma_{KN}}{d\Omega'_1} = \frac{3 \sigma_T}{16 \pi} \left( \frac{\nu'_1}{\nu'} \right)^2 \left[ \frac{\nu'}{\nu'_1} + \frac{\nu'_1}{\nu'} - \sin^2 (\theta'_1 - \theta')  \right],   
\end{equation}
and, again, one can use equation (\ref{eq:IC}) to eliminate $\nu'_1$, 
which yields


\begin{equation} \label{eq:KN}
\frac{d\sigma_{KN}}{d\Omega'_1} = \frac{3 \sigma_T}{16 \pi} \left[ \frac{1}{1+x'[1 - \cos (\theta'_1 - \theta')]}  \right]^2 
\left[ \frac{1}{1+x'[1 - \cos (\theta'_1 - \theta')]} + x'[1 - \cos (\theta'_1 - \theta')] - \cos^2 (\theta'_1 - \theta') \right].   
\end{equation}

Equation (\ref{eq:generaldEdt}) is general and contains no 
assumptions.  To simplify the calculation we now make  
two assumptions.  The first assumption allows us to simplify 
the expression of $\cos (\theta'_1 - \theta')$ the following way.  
The integrand of $d\Omega'_1$ in equation (\ref{eq:generaldEdt}) 
depends on $\theta'$, and both $x'$ and $d\sigma_{KN}/d\Omega'_1$ 
also depend on $\theta'$ -- see definition of $x'$ and equations 
(\ref{eq:Dopler1}) and (\ref{eq:KN}).  The cross
section starts to fall-off steeply with angle only 
when $|\theta'_1 - \theta'| \gae 1/\sqrt{x'}$; and 
also $\theta' \approx \gamma_e^{-1}$ -- see equation (\ref{eq:Dopler2}).
Thus, for parameters of interest to us, where $\gamma_e \gae 10^3$
and $x' \lae 10^2$, we have $\theta'_1 \sim 0.1$ and 
$\theta' \sim 10^{-3}$.  In this case, we can approximate
$\cos (\theta'_1 - \theta') \approx \cos \theta'_1$.  We call 
this the ``head-on'' approximation, since it corresponds to an 
incoming photon moving in the same direction as the incoming 
electron velocity vector before the collision as seen in the
electron rest frame.  We note, however, that we cannot set 
$\theta'=0$ in equation (\ref{eq:Dopler1}), since that would 
overestimate $\nu'$ by a factor of $\sim 2$.  The second assumption 
we make is that the energy density in photons in the co-moving 
frame of the source where the photons are generated 
is distributed isotropically, therefore, we can write

\begin{equation}
d\Omega' d\nu' I'_{\nu'} = d\Omega d\nu I_{\nu} \left( \frac{\nu'}{\nu} \right)^2 = d\Omega d\nu \frac{u_{\nu}}{4 \pi} c \gamma_e^2 (1 + \beta_e \cos \theta )^2,
\end{equation}
where $u_{\nu}$ is the source rest frame photon energy density.  To derive the last 
expression we have used the fact that $I_{\nu}/\nu^3$ is a Lorentz invariant 
quantity and that $d\Omega'=d\Omega(\nu/\nu')^2$.  With these two 
assumptions we can rewrite equation (\ref{eq:generaldEdt}) as

\begin{equation} \label{eq:dEdtassumptions}
\frac{dE_s}{dt} = \frac{c \gamma_e^2}{2} \int_0^{\infty} \! d\nu u_{\nu} \int_{-1}^{1} \! d\mu (1 + \beta_e \mu )^2 \int_{-1}^{1} \! d\mu'_1  \frac{1 - \beta_e \mu'_1}{1+x'(1 - \mu'_1)} \frac{d\sigma_{KN}}{d\mu'_1},   
\end{equation}
and the Klein-Nishina cross section is given by

\begin{equation} \label{eq:KNassumptions}
\frac{d\sigma_{KN}}{d\mu'_1} = \frac{3 \sigma_T}{8} \frac{1}{[1+x'(1 - \mu'_1)]^2} \left[ \frac{1}{1+x'(1 - \mu'_1)} + x'(1 - \mu'_1) - {\mu'_1}^2 \right],   
\end{equation}
where $\mu = \cos \theta$, $\mu'_1=\cos \theta'_1$, 
and, as defined before, $x' = h \gamma_e \nu (1 + \beta_e \mu)/m_e c^2$.
Combining equations (\ref{eq:dEdtassumptions}) and (\ref{eq:KNassumptions})
we find


\begin{equation} \label{eq:dEdtfinal}
\frac{dE_s}{dt} = \frac{3}{16} \sigma_T c \gamma_e^2 \int_0^{\infty} \! d\nu u_{\nu} \int_{-1}^{1} \! d\mu (1 + \mu )^2 
\int_{-1}^{1} \! d\mu'_1 \left[ \frac{1-\mu'_1}{[1+x'(1 - \mu'_1)]^4} + \frac{x'(1-\mu'_1)^2}{[1+x'(1 - \mu'_1)]^3} - 
\frac{{\mu'_1}^2 (1-\mu'_1)}{[1+x'(1 - \mu'_1)]^3} \right], 
\end{equation} 
where we made another approximation, which is that $\beta_e \approx 1$, 
which is valid for the case, where $\gamma_e \gg 1$ studied here.

The $\mu'_1$ integral in equation (\ref{eq:dEdtfinal}) can be 
carried out analytically.  Let us define a function

\begin{equation}
G_n(x) = \int_{-1}^{1} \! \frac{d\mu'_1}{[1+x(1-\mu'_1)]^n}.
\end{equation}
It is straightforward to show that 

\begin{eqnarray}
G_1(x) & = & \frac{\ln(1+2x)}{x}, \nonumber \\
G_n(x) & = & \frac{1}{(n-1)x} \left[1 - \frac{1}{(1+2x)^{n-1}}\right], \qquad \textrm{for } n \ne 1. 
\end{eqnarray}
With this last equation, we can evaluate the $\mu'_1$ integral
of the first term in equation (\ref{eq:dEdtfinal}), 
which is

\begin{equation} \label{eq:integrand1}
\int_{-1}^{1} \! d\mu'_1 \frac{1-\mu'_1}{[1+x'(1-\mu'_1)]^4} = -\frac{1}{3}\frac{d}{dx'}G_3(x')= \frac{2}{3}\frac{(3+2x')}{(1+2x')^3},
\end{equation}
and also the second term, which yields

\begin{equation} \label{eq:integrand2}
\int_{-1}^{1} \! d\mu'_1 \frac{x'(1-\mu'_1)^2}{[1+x'(1-\mu'_1)]^3} = \frac{x'}{2}\frac{d^2}{d{x'}^2}G_1(x')= \frac{\ln(1+2x')}{{x'}^2} - \frac{2(1+3x')}{{x'}^2(1+2x')^2}
\end{equation}
To carry out the integral of the third term, we define a new function 

\begin{equation}
J_n(x) = \int_{-1}^{1} \! d\mu'_1 \frac{\mu'_1}{[1+x(1-\mu'_1)]^n},
\end{equation}
which can be shown to be

\begin{eqnarray} \label{eq:Jfunc}
J_1(x) & = & \frac{x+1}{x^2}\ln(1+2x) - \frac{2}{x}, \nonumber \\
J_n(x) & = & \frac{1}{(n-1)x} \left[1 + \frac{1}{(1+2x)^{n-1}}\right] - \frac{1}{(n-1)x} G_{n-1}(x), \qquad \textrm{for } n \ne 1. 
\end{eqnarray}

Using partial fractions decomposition, the $\mu'_1$ integral 
of the third term in equation 
(\ref{eq:dEdtfinal}) can be written as

\begin{equation}
\int_{-1}^{1} \! d\mu'_1 \frac{{\mu'_1}^2 (1-\mu'_1)}{[1+x'(1-\mu'_1)]^3} = - \frac{1}{{x'}^2}J_1(x') + \frac{2+x'}{{x'}^2}J_2(x') - \frac{1+x'}{{x'}^2}J_3(x'),
\end{equation}
and by using equation (\ref{eq:Jfunc}), we find 

\begin{equation}  \label{eq:integrand3}
\int_{-1}^{1} \! d\mu'_1 \frac{{\mu'_1}^2 (1-\mu'_1)}{[1+x'(1-\mu'_1)]^3} = - \frac{2x'+3}{{x'}^4}\ln(1+2x') + \frac{2{x'}^3 +20{x'}^2+22x'+6}{{x'}^3(1+2x')^3}.
\end{equation}
Combining equations (\ref{eq:integrand1}), (\ref{eq:integrand2}) and 
(\ref{eq:integrand3}), the $\mu'_1$ integral in equation 
(\ref{eq:dEdtfinal}) can be written as 

\begin{equation}
K(x') = \frac{{x'}^2 + 2x' + 3}{ {x'}^4 }\ln(1+2x') - \frac{2 ( 22 {x'}^4 + 75 {x'}^3  + 99 {x'}^2 + 51 x' + 9 ) }{3 {x'}^3 (1+2x')^3}.
\end{equation}
In addition, we make one last transformation of equation 
(\ref{eq:dEdtfinal}) and that is the following:  Since 
$x'$ depends on $\mu=\cos \theta$, we write it as

\begin{equation}
x' = \frac{h \nu \gamma_e}{m_e c^2}(1+\beta_e \cos \theta) \approx X(1+\mu),
\end{equation}
where we have defined a new variable $X = h \nu \gamma_e/m_e c^2$, 
and we again have made use of the assumption that $\gamma_e \gg 1$, so that 
$\beta_e \approx 1$. With this, equation (\ref{eq:dEdtfinal})
can be rewritten as

\begin{equation} \label{eq:dEdtfinal2}
\frac{dE_s}{dt} = \frac{3}{16} \sigma_T c \gamma_e^2 \int_0^{\infty} \! d\nu \frac{u_{\nu}}{X^3} \int_{X(1-\beta_e)}^{X(1+\beta_e)} \! dx' {x'}^2 K(x'). 
\end{equation}
The $x'$ integral can be done analytically, except for one term in the 
integrand, which is $2\ln(1+2x')/x'$, whose integral is a polylogarithm, 
specifically a dilogarithm or Spence's function.

The rate of Inverse Compton energy loss can be expressed as 

\begin{equation}
\frac{d}{dt}(m_e c^2 \gamma_e) = - \frac{dE_s}{dt} + \int \! d\nu \; u_{\nu} \sigma_{KN} c \approx - \frac{dE_s}{dt}
\end{equation}
where $dE_s/dt$ is given by equation (\ref{eq:dEdtfinal2}), 
and the second term is the integral of photon energy before scattering which 
is negligible compared with $dE_s/dt$ for $\gamma_e \gg 1$.  

Finally, the Compton-$Y$ parameter, defined as the ratio of the 
rate of Inverse Compton energy loss 
and the rate of synchrotron energy loss, is given by 

\begin{equation} \label{eq:Y}
Y(\gamma_e) = \frac{\frac{dE_s}{dt}}{\frac{1}{6 \pi} \sigma_T B^2 \gamma_e^2 c}.
\end{equation}

The only thing left to determine now is the energy density in photons, 
$u_{\nu}$.  Since in the case of GRBs the entire source (jet) is moving 
relativistically towards the observer, then the photon energy density 
needed is the one measured in the source co-moving frame.   
All quantities considered in the following calculation will 
be in the source co-moving frame.

The power emitted in $4 \pi$ sr at 
frequency $\nu$ by an electron in the shell 
(in units of erg s$^{-1}$ sr$^{-1}$ Hz$^{-1}$) is (see, e.g., 
RL79)

\begin{equation} \label{eq:Pnu}
P_{\nu} = \frac{e^3 B}{m_e c^2}\left(\frac{\nu}{\nu_{\gamma_e}}  \right)^{1/3},
\end{equation}
where $e$ is the electron charge, $B$ is the magnetic field in the shell
co-moving frame and $\nu_{\gamma_e}$ is the synchrotron frequency 
of an electron with LF $\gamma_e$, which is given by 

\begin{equation} \label{eq:nu2}
\nu_{\gamma_e} = \frac{e B \gamma_e^2}{2 \pi m_e c}.
\end{equation}
The numerical factors in the last two expressions are  
different than the ones in  Wijers \& Galama (1999) 
for $2 \lae p \lae 3$ by only less than $\sim 10$\% and $\sim 40$\% for 
equations (\ref{eq:Pnu}) and (\ref{eq:nu2}), respectively.
$P_{\nu}$ in equation (\ref{eq:Pnu}) is valid for $\nu < \nu_{\gamma_e}$,
for $\nu > \nu_{\gamma_e}$ we take $P_{\nu}$ to vanish, even though 
strictly speaking it decreases exponentially, which introduces a very 
small error.

\begin{figure}
\begin{center}
\includegraphics[width=8cm, angle=0]{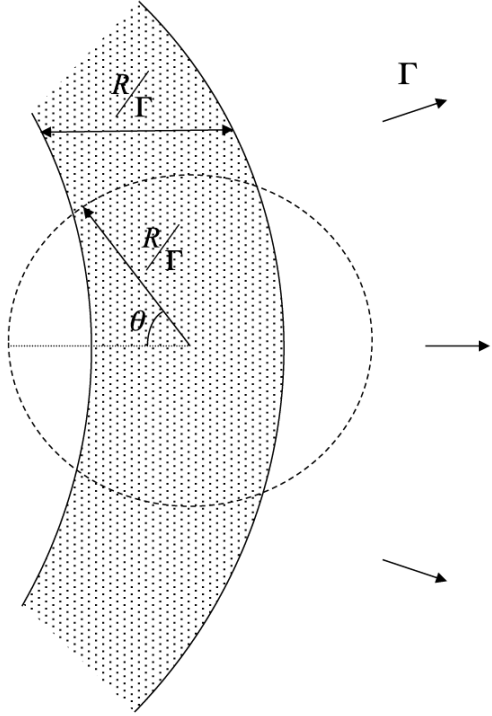}
\end{center}
\caption{ Schematic of a GRB jet.  The width of the 
shell in the co-moving frame of the shell is $R/\Gamma$, 
where $R$ is the distance from the center of the explosion 
and $\Gamma$ is the LF of the source.  To calculate
the specific intensity in the middle of the shell 
one needs to take into account all photons produced by 
electrons (emitters) within the causally 
connected region of radius $R/\Gamma$. Since 
$R/\Gamma \ll R$ we assume a rectangular slab geometry.}
\label{fig2} 
\end{figure}

We will now calculate the specific intensity, $I_{\nu}$, 
in the middle of the shell (see Figure \ref{fig2}).  
Let us assume that the 
column density of electrons (emitters) in the shell is $N$
(number of emitters per unit area) and the shell 
radius and LF are $R$ and $\Gamma$, respectively.
Since the photons cannot arrive at a point in the middle 
of the shell from a distance larger than $R/\Gamma$, which is the 
radius of the causally connected region, then the specific 
intensity in the middle of the shell is approximately
given by

\begin{equation} 
I_{\nu}(\theta) =  \left\{ \begin{array}{lll}
\frac{P_{\nu}}{4 \pi} \frac{N}{2}|\sec \theta| & \textrm{for $0 \le \theta \le \pi/3$ and $2\pi/3 \le \theta \le \pi$} \\
\\
\frac{P_{\nu} N}{4 \pi} & \textrm{for $\pi/3 < \theta < 2\pi/3$}, \\
\end{array} \right.
\end{equation}
where we have used a rectangular slab geometry 
since $R/\Gamma \ll R$.
Therefore, we find that the energy density in photons 
of frequency $\nu$ in the middle of the shell is 

\begin{equation}
u_{\nu} = \frac{1}{c} \int \! d\Omega I_{\nu} (\theta) = \frac{P_{\nu} N}{c}\left(\frac{\ln 2}{2} + \frac{1}{2}\right).
\end{equation}
A similar analysis shows that the photon energy 
density near the inner or outer edge of the shell
is $u_{\nu} = P_{\nu} N/2 c$.  Thus, the average 
value of $u_{\nu}$ in the shell is $u_{\nu} \approx 0.7 P_{\nu} N/c$.

The column density of electrons can be written as a function of 
the electron energy distribution, which is defined as 
$n(\gamma_e)$ (number of electrons per unit area per unit 
$\gamma_e$).  We define the specific flux in the co-moving frame 
of the shell, $f_{\nu} \equiv P_{\nu} N$, thus, $u_{\nu} \approx 0.7 f_{\nu}/c$.
The specific flux is given by

\begin{equation}
f_{\nu} = \int \! d\gamma_e n(\gamma_e) P_{\nu}, 
\end{equation}
where $n(\gamma_e)$ is calculated self-consistently 
by solving a coupled set of equations for $n(\gamma_e)$
and radiation (see below).  Using equation (\ref{eq:Pnu}) 
we rewrite the last equation as

\begin{equation} \label{eq:syn}
f_{\nu} = \int_{\gamma_{\nu}}^{\infty} \! d\gamma_e n(\gamma_e) 
\left[ \frac{e^3 B}{m_e c^2} \right] \left( \frac{\gamma_{\nu}}{\gamma_e}\right)^{2/3}, 
\end{equation}
where $\gamma_{\nu}$ is the 
LF of electrons radiating at synchrotron frequency $\nu$, that is
-- see equation (\ref{eq:nu2}) --, 

\begin{equation}
\gamma_{\nu}^2 = \frac{2 \pi m_e c \nu}{e B}.
\end{equation}

The task now is to determine the electron energy distribution, $n(\gamma_e),$
in order to calculate the synchrotron flux given by equation
(\ref{eq:syn}).  The electron distribution function is determined by solving 
the following continuity equation 

\begin{equation} \label{eq:continuity}
\frac{\partial n(\gamma_e)}{\partial t} + \frac{\partial}{\partial \gamma_e} 
\left[ \dot{\gamma_e} n(\gamma_e) \right] = S(\gamma_e),
\end{equation}
where the source term, $S(\gamma_e)$, is given by 

\begin{equation} 
S(\gamma_e) = \left\{ \begin{array}{ll}
\dot{N} \left(\frac{\gamma_e}{\gamma_i}  \right)^{-p}  & \textrm{$\gamma_e \ge \gamma_i$} \\
0 & \textrm{$\gamma_e < \gamma_i$}, \\
\end{array} \right.
\end{equation}
$\dot{N}$ is the total number of electrons crossing 
the shock front per unit time per unit area, and $\gamma_i$
is the minimum LF of electrons injected in the shock.

We can solve equation (\ref{eq:continuity}) approximately the following way.  Let us 
define a cooling time for electrons,
$t'_{cool}$, in the shell co-moving frame, as follows
$t'_{cool}(\gamma_e)=\gamma_e/\dot{\gamma_e}$ and a co-moving 
dynamical time, $t'_{co}$, as $t'_{co}=R/(\Gamma c)$.   
The dynamical time in the shell co-moving frame is related to the 
one in the observer frame, $t_{obs}$, as  $t'_{co} = \Gamma t_{obs}/(1+z)$, 
where $z$ is the redshift and the column density is related to the 
shock radius, $R$, as $nR/3$, therefore, $\dot{N}=nR/3t'_{co}$.

If $t'_{cool} > t'_{co}$ for a given $\gamma_e$, 
electrons of this LF have not cooled much in the available time, 
therefore, equation (\ref{eq:continuity}) reads

\begin{equation} 
n(\gamma_e) \approx t'_{co} S(\gamma_e), \qquad \textrm{for } t'_{cool} > t'_{co}. 
\end{equation}
On the other hand, when $t'_{cool} < t'_{co}$ for a given $\gamma_e$,
then $n(\gamma_e)$ wouldn't change with time 
and thus

\begin{equation} 
\frac{\partial}{\partial \gamma_e} \left[ \dot{\gamma_e} n(\gamma_e) \right] \approx S(\gamma_e).
\end{equation}
or

\begin{equation} \label{eq:IClast}
n(\gamma_e) = \int_{\gamma_e}^{\infty} \! d\gamma_e'S(\gamma_e')\bigg/\dot{\gamma_e}, \qquad \textrm{for } t'_{cool} < t'_{co}.
\end{equation}

To summarize, we solve numerically the following equations {\it simultaneously}
to find Compton $Y(\gamma_e)$: equation (\ref{eq:continuity}) to 
find the electron energy distribution, equation (\ref{eq:syn}) to 
find the synchrotron flux, equation (\ref{eq:dEdtfinal2}) to find 
the energy loss rate due to Inverse Compton and substituting all these 
results into equation (\ref{eq:Y}) gives $Y(\gamma_e)$.

\section{Solution for GRB 090902B}

We use the parameters of GRB 090902B, $z=1.8$ (Cucchiara
et al. 2009), luminosity distance of 
$d_L=4.3\times10^{28}$ cm and we also use $p=2.2$ (see Section 3.1) and 
investigate Option 2 ($\nu_i < \nu_{opt} < \nu_c < \nu_x$) in detail.
For this, we use the formalism developed in the previous section.

We use a numerical code that scans the 4 external 
forward shock parameters, viz. $\epsilon_e$, $\epsilon_B$, $n$ and $E_{KE,iso}$,
and finds the subset that satisfies a set of chosen constraints (see below).
The parameters are varied in the following ranges:
$\epsilon_e=10^{-2}-10^{-0.2}$, $\epsilon_B=10^{-9}-10^{-1}$,
$n=10^{-4}-10$ cm$^{-3}$ and $E_{KE,iso} = 10^{53}-10^{55.5}$ erg.
The value of the isotropic energy released in gamma-rays 
for GRB 090902B was $E_{\gamma,iso}=3.63\times10^{54}$ erg
(Abdo et al. 2009), this is the reason why we chose to vary $E_{KE,iso}$ from being 
$30$ times smaller to $\sim 10$ times larger than $E_{\gamma,iso}$.
The radiation efficiency of the prompt gamma-ray 
emission is given by 
$\eta = E_{\gamma,iso}/(E_{\gamma,iso} + E_{KE,iso})$.  This efficiency 
will be useful when trying to further constrain our results.

We numerically calculate the allowed subspace of 4-D parameter space for 
GRB 090902B by imposing {\it only two} constraints: (1) the theoretically calculated 
X-ray flux at 1 keV at $12.5$ h, the first X-ray data, must be consistent 
within 1-$\sigma$ with the observed value of 
$0.38 \pm 0.10$ $\mu$Jy (Cenko et al. 2011, Pandey et al. 2010) and (2) the 
injection frequency ($\nu_i$) at this same time should be below the optical band ($\nu_i < 2$ eV)
so that the optical light curve decays with time for $t > 12.5$ h. 
 
\begin{figure}
\begin{center} 
\includegraphics[width=11cm, angle=0]{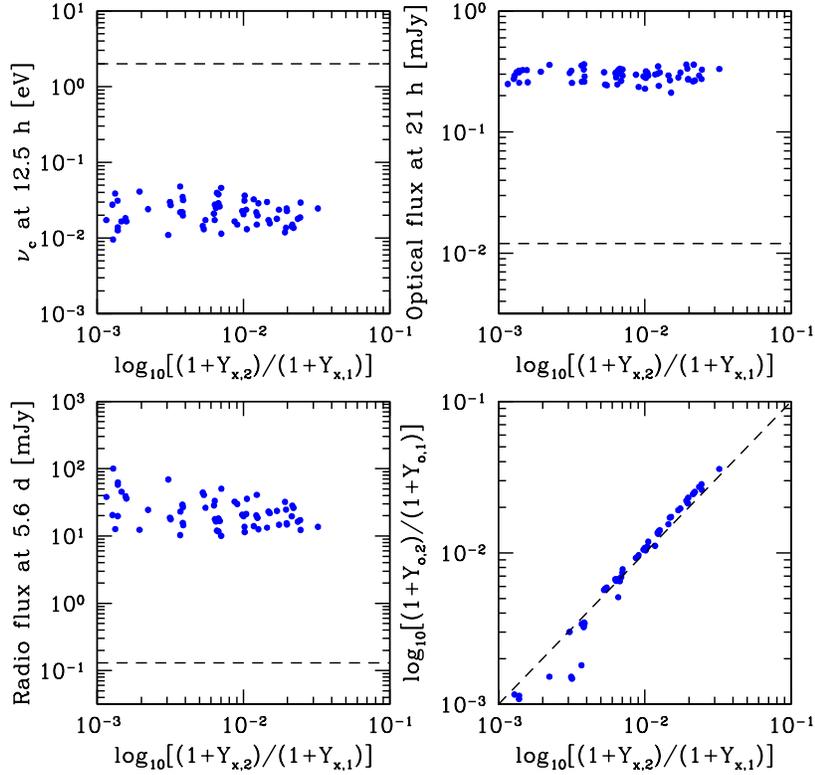}
\end{center}
\caption{We solve for the allowed 4-D external forward shock 
parameter space for constant CSM and $p=2.2$ ($\epsilon_e$, $\epsilon_B$, $n$, $E_{KE,iso}$) 
by imposing 
only two constraints: (i) The X-ray flux at $t_1=12.5$ h should 
be consistent with the observed value and (ii) the injection 
frequency, $\nu_i$, should be $< 2$ eV at the same time
(see text). For this allowed parameter space we calculate (blue points)
the cooling frequency at $t_1$ (top left), the optical 
flux at 21 h (top right), the radio flux at 5.6 d (bottom left) 
and the Compton-$Y$ for electrons radiating at 2 eV, $Y_o$
(bottom right).  All quantities are plotted as a function of 
the ratio of Compton-$Y$ for electrons radiating at 1 keV, $Y_x$, 
at two times, $t_1$ and $t_2 = 10 t_1$, with subscripts 
``1'' and ``2'', respectively.  We {\it only} plot the data for which 
$(1+Y_x)$ increases with time.  In order to steepen the X-ray 
light curve for GRB 090902B, with $\nu_i<\nu_{opt}<\nu_c<\nu_x$ 
(Option 2), to make it consistent with the observed behavior requires
$\log_{10} [(1+Y_{x,2})/(1+Y_{x,1})] = 0.21$, which is not 
found for any point in the 4-D parameter space. Also, 
we find $\nu_c<2$ eV, inconsistent with 
Option 2 (the horizontal dashed line shows $\nu_c=2$ eV
-- top left). Moreover, the optical and radio 
fluxes are inconsistent with the observed values 
(horizontal dashed lines).  The bottom 
right panel shows that $(1+Y_x) \propto (1+Y_o)$, and therefore whenever
the X-ray light curve is steepened due to the increase of $Y_x$ with time, 
the optical light curve is also steepened by the same amount.}
\label{fig3} 
\end{figure}

For the resulting subspace of 4-D parameter space we calculate a few
quantities of interest for this GRB.  (1) Since we want to 
address the possibility of the X-ray decay being faster
due to a time dependence of the Compton-$Y$
parameter as $(1+ Y_x) \propto t^{0.21}$ (for Option 2: $\nu_i < \nu_{opt} < \nu_c < \nu_x$), 
we calculate $Y_x$ for 
electrons radiating at 1 keV at two times, $t_1=12.5$ h and 
$t_2=10 t_2$, which we define as $Y_{x,1}$ and $Y_{x,2}$, respectively.  
These times span almost the entire duration 
of the X-ray observations for GRB 090902B.  Steepening the X-ray light curve
to the observed value would require 
$\log_{10} [(1+Y_{x,2})/(1+Y_{x,1})] = 0.21$.  (2) We 
calculate the synchrotron cooling frequency 
at $t_1$ to see if it is consistent with the 
orderings considered above, that is, $\nu_{opt} < \nu_c < \nu_x$. (3) We calculate 
the optical flux at 21 h, which we compare to the observed 
value of $12.0 \pm 0.1$ $\mu$Jy (Pandey et al. 2010).  (4) The radio flux at 
8.46 GHz is also calculated at 5.6 d, and it is compared to the 
observed value of $130 \pm 34$ $\mu$Jy detected by the Very Large Array 
(Cenko et al. 2011). (5) We 
calculate the Compton-$Y$ parameter for electrons radiating 
at $\lae 2$ eV (optical), $Y_o$, at 
$t_1$ and $t_2$, to see if $Y_o$ and $Y_x$ behave differently or not. 
The results of our numerical calculation are shown in Figure
\ref{fig3}.  

As can be seen from Figure \ref{fig3}, no part of the parameter 
space reaches the desired value of 
$\log_{10} [(1+Y_{x,2})/(1+Y_{x,1})]= 0.21$. The maximum 
value reached is $\log_{10} [(1+Y_{x,2})/(1+Y_{x,1})] \sim 0.03$, 
which means that the X-ray decay would only steepen at 
most to $t^{-1.18}$, which is inconsistent with the observed 
value by more than 5-$\sigma$.  Moreover,  
the optical and radio fluxes are inconsistent with the
observed values by a factor of $\sim 20$ and $\sim 100$, respectively (Fig. \ref{fig3}).
More importantly, we also find that $\nu_c$ at $12.5$ h is 
actually below the optical band, which is inconsistent with the frequency
ordering we are considering here (see Fig. \ref{fig3}).  All these issues rule out the possibility that 
the temporal evolution of Compton-$Y$ can explain the observed X-ray data.

In Option 1, the optical and X-ray bands are below $\nu_c$, and thus 
the fluxes in these bands do not depend on Compton-$Y$ -- see equation (\ref{eq:flux}).  For this 
reason the temporal evolution of Compton-$Y$ cannot be invoked 
in this case to reconcile the difference between the data and the expectation 
of the external forward shock model.

Let us now explore Option 3, where both optical and X-ray bands lie 
above the cooling frequency, therefore the fluxes in both 
bands in this case depend on Compton-$Y$.  For $p=1.5$, the optical 
flux would decay as $\propto t^{-(3p+10)/16}\nu^{-p/2}=t^{-0.91}\nu^{-0.75}$ and both the 
optical spectrum and optical decay index are consistent with the 
observed values within 1-$\sigma$.  The X-ray decay then, 
must be steepened by $t^{-0.45}$ to match the observed value, therefore, 
we check to see if $(1+Y_x)\propto t^{0.45}$ is allowed by a subset of the 
4-D parameter space. Moreover, $(1+Y_o)$ should evolve 
very slowly with time, otherwise the optical light curve would 
{\it also} steepen and that would be inconsistent with the observed optical 
data.  

Our numerical code is unable to handle $p<2$, thus 
we cannot calculate the allowed parameter space for $p=1.5$.  However, 
we can run our code for $p = 2.05$ and compare our results 
with $p=2.2$.  We impose the same two observational constraints as before.
For $p=2.05$  we find that the magnitude of $\log_{10} [(1+Y_{x,2})/(1+Y_{x,1})]$ remains 
roughly the same as it is for $p=2.2$; however, $\log_{10} [(1+Y_{o,2})/(1+Y_{o,1})]$
increases by a constant factor of $\sim 2$.  This means that 
$(1+Y_o)$ increases with time faster than $(1+Y_x)$.
We expect this behavior to continue when we decrease the 
value of $p$ to $p=1.5$ (there is no reason for an abrupt change when $p$
falls below 2) therefore, the effect of Inverse Compton cooling on the electron energy distribution 
in this case is to steepen the optical light curve more than the X-ray light curve. 
This suggests that the temporal evolution of Compton-$Y$ in Option 3 to explain the observed data can be 
also ruled out. 

\subsection{What is the real solution?}

If the X-ray, optical and radio data are consistent with the 
external forward shock model predictions, then we should be able to 
find a subset of the 4-D parameter space for which the 
optical and radio fluxes agree with the observed values
(the X-ray flux agrees with the observed value by design).   
Although we just found out that this subspace does not exist 
when we require $(1+Y_x)$ to increase with time (see Figure \ref{fig3}), 
this subspace does exist when $(1+Y_x)$ {\it decreases} with time (not plotted 
in Figure \ref{fig3}).  

\begin{figure}
\begin{center}
\includegraphics[width=12cm, angle = 0, clip=true, viewport=.0in .0in 8in 4.5in]{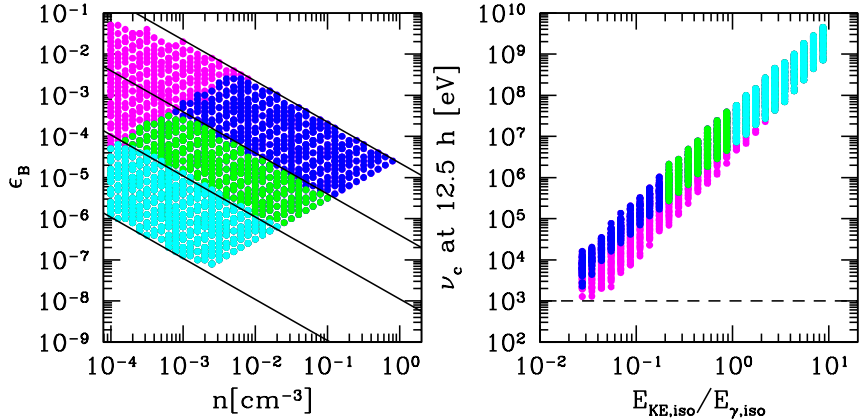}
\end{center}
\caption{Allowed $\epsilon_B$-$n$ plane and $\nu_c$ at 12.5 h as a function of 
$E_{KE,iso}/E_{\gamma,iso}$ (left and right panels, respectively) 
for constant CSM and $p=2.2$ when the X-ray, optical and radio 
fluxes predicted by the external forward shock model at 12.5 h, 21 h and 5.6 d, respectively, 
are consistent with the observed values within 1-$\sigma$ (Constraint 1).  We further narrow down
the allowed parameter space with the following constraints: $\epsilon_e>0.2$
(Constraint 2 - see Section 5), $E_{KE,iso}/E_{\gamma,iso}>0.2$ (Constraint 3) 
and $E_{KE,iso}/E_{\gamma,iso}>1$ (Constraint 4).  The points 
are color coded according to the applied constraints 
(in parenthesis): magenta (1), blue (1 and 2), green (1, 2 and 3)
and cyan (1, 2 and 4).  The solid black lines show the expectations 
for the shock-compressed magnetic field of seed values 
1, 10, 60 and 450 $\mu$G (from bottom to top -- see text). The horizontal 
dashed line shows $\nu_c=1$ keV.  The injection frequency at 12.5 h is 
$\nu_i \sim 0.03$ eV. Notice that these figures have many more points 
than the ones on Figure \ref{fig3}, since there are many more solutions for 
which $(1+Y_x)$ decreases with time.  For the points in these figures, we 
find that, at most, $(1+Y_x)\propto t^{-0.1}$, however, as seen in the 
right panel, $\nu_x < \nu_c$, therefore, the X-ray temporal decay index is 
unaffected. (Please, see online version for color figure.)}
\label{fig4} 
\end{figure}

We take now the result of the parameter search in the previous 
subsection and further 
constrain it with the following conditions: (i) The external 
forward shock optical flux at 21 h should be consistent with 
the observed value within 1-$\sigma$, and (ii) the external 
forward shock radio flux at 5.6 d should be consistent with 
the observed value within 1-$\sigma$.  This gives us a 
subspace of 4-D parameter space for which the X-ray, optical and radio 
fluxes as predicted by the synchrotron process in the 
external forward shock agree with the observed values 
within 1-$\sigma$. We show this subspace projected on 
the $\epsilon_B$-$n$ plane in Figure \ref{fig4} 
and we show $\nu_c$ at 12.5 h as a function of $E_{KE,iso}/E_{\gamma,iso}$.  
We compare our result of $\epsilon_B$ with the expectation of a 
magnetic field that is shock-compressed CSM 
field with pre-shocked value of $B_0$. The value of
$\epsilon_B$ downstream of the shock-front 
resulting from the shock compressed CSM field is 
$\epsilon_B\approx {B_0}^2/(2\pi n m_p c^2)$, where $m_p$ 
is the proton mass and $n m_p$ is the CSM mass density.
As shown in Fig. \ref{fig4}, $B_0 \sim 10$ $\mu$G can explain 
all the afterglow radiation without the need for a 
strong dynamo amplification of shock compressed field.  

We find that $\nu_c$ at 12.5 h is always $\gae 1$ keV. Note 
that $\nu_c \sim 1$ keV -- within a factor of $\sim 3$ -- 
only for 
$E_{KE,iso}/E_{\gamma,iso} \lae 0.06$, which would 
require an extremely high efficiency of $\gae 95$\% in producing 
the prompt gamma-rays.  The overwhelming majority 
of the parameter space agrees very well with the expectations of Option 1, 
that is, $\nu_c$ is above 1 keV at 12.5 h.  
However, how do we reconcile the different 
temporal decay indices of the X-ray and optical 
light curves? This is addressed below.

\subsection{Curvature in the injected electron spectrum}

The only option left to explore is the possibility that 
the spectrum of injected electrons exhibits 
some curvature, that is, that the value of $p$ is 
not the same for all observed bands, but that it is a 
function of electron energy.  In this scenario, we allow 
$p$ to vary and determine the best $p$ values -- the ones that give us 
the best agreement within the observed uncertainties -- which are consistent with: (i) the
observed optical spectrum and temporal decay ($p_{opt}$), and (ii) consistent with the 
observed X-ray spectrum and temporal decay ($p_x$).

We will consider the only viable option we have found --
$\nu_i < \nu_{opt} < \nu_x < \nu_c$ (Option 1) -- 
and calculate the required values of  $p$ from the data.  From the observed 
$\alpha_{opt}$ and $\beta_{opt}$, we find that $p_{opt} = 2.2$ would consistently explain the optical data 
to within $\sim2$-$\sigma$, and for the observed values of $\alpha_x$ and $\beta_x$ 
we find that $p_x=2.8$ would consistently explain the optical data to within
1-$\sigma$.

If $\nu_i < \nu_{opt} < \nu_c < \nu_x$ (Option 2), then 
the values of $p_{opt}$ and $p_x$ that best fit the  
data are $p_{opt}=2.2$ (to within $\sim2$-$\sigma$) and $p_x=2.5$ (to 
within 3-$\sigma$). On the other hand, if $\nu_i < \nu_c < \nu_{opt} < \nu_x$ 
(Option 3), then the values of $p_{opt}$ and $p_x$ that best fit 
the data are $p_{opt}=1.4$ (to within 1-$\sigma$) and $p_x=2.5$ (to 
within 3-$\sigma$). Requiring a change in $p$ in Option 
3 from $p_{opt}=1.4$ to $p_x=2.5$ seems unlikely, since this change 
is very similar to the change one expects due to the cooling frequency, which 
for Option 3 should be below the optical band.  Moreover, the X-ray data
only agrees within 3-$\sigma$.  For these two reasons, we rule out Option 3.
It might seem that Option 2 provides a good solution, since $p_{opt} \approx p_x$, 
however, according to Figure \ref{fig4}, for $p=2.2$ ($p_{opt}=p_x$) there 
is no parameter space where $\nu_c < \nu_x$.  If we repeat the calculation 
for $p=2.35$, which is an average of $p_{opt}$ and $p_x$ in Option 2, 
we find that Figure \ref{fig4} is basically unchanged.  For this reason, we rule out 
the possibility that Option 2 with a curvature in the electron distribution
function can explain the observed data.  Therefore, Option 1 is the only 
viable solution for the afterglow data of GRB 090902B.     

In conclusion, we find that the external forward shock model can explain 
the  afterglow data for GRB 090902B provided that the 
cooling frequency ($\nu_c$) is larger than 1 keV at $\sim 0.5$ d 
and $\nu_i < 2$ eV.  We also find that in order 
to explain the different temporal decay indices of the optical 
and X-ray light curves, there must be a slight curvature in the 
electron energy distribution function, where the spectrum of injected electrons steepens from 
$\propto \gamma_e^{-2.2}$ to $\propto \gamma_e^{-2.8}$, when $\gamma_e$
increases by a factor of $\sim 30$, corresponding to electron synchrotron frequency
increasing from optical to $\sim 1$ keV.  This happens
effectively at a LF which corresponds to synchrotron frequency $\nu_b$. 
The energy spectrum below (above)
$\nu_b$ should be $\propto \nu^{-0.60}$ ($\propto \nu^{-0.90}$) and the light 
curves should decay as $\propto t^{-0.90}$ ($\propto t^{-1.35}$). In the 
next section we determine this break frequency ($\nu_b$), although, one should 
keep in mind that this break might not be sudden but is probably gradual. 

\subsubsection{Break frequency}

Using the X-ray and optical data at a specific time, we can 
determine the effective break frequency necessary to 
explain the observations.  
At 21 h, the optical flux (2 eV) is $12$ $\mu$Jy and the X-ray flux 
(at 1 keV) is $0.2$ $\mu$Jy (Pandey et al. 2010).  At this time, one can 
show that a single power-law spectrum with $\nu^{-0.60}$ reconciles 
these two fluxes, which means that $\nu_b$ should be very close to 
1 keV. We estimate $\nu_b$ at 21 h by assuming that the 
spectrum below (above) $\nu_b$ is $\nu^{-0.60}$ ($\nu^{-0.9}$)
and find that $\nu_b$ should be 0.3 keV (this validates the calculation 
of the parameter space shown in Figure \ref{fig4}:
at 21 h the curvature of the spectrum is just starting to 
be evident in the X-ray data).  One expects $\nu_b$
to decrease with time, therefore, the evidence for the proposed 
curvature in the spectrum should become stronger with time.

We use the combined spectrum presented in Pandey et al. (2010)
fig. 2 at 1.9 d to determine the location of $\nu_b$ at later times.
The optical flux (2 eV) at this time is $7$ $\mu$Jy and the X-ray flux 
(at 2.88 keV) is $0.03$ $\mu$Jy.  Again, assuming that the 
spectrum below (above) $\nu_b$ is $\nu^{-0.60}$ ($\nu^{-0.9}$), 
we can determine $\nu_b$ to be $\approx 80$ eV. 
The values of $\nu_b$ at 21 h and 1.9 d
allow us to find that it decreases with time as $\sim t^{-1.75}$, 
which is roughly the same time dependence as $\nu_i$, which exhibits $\nu_i \propto t^{-1.5}$.  

At the beginning of the {\it Swift} XRT observations 
at 12.5 h, we therefore expect $\nu_b$ to be $\sim 0.8$ keV, 
and at 21 h, we expect $\nu_b$ to be $\sim 0.3$ keV, which is the 
lower bound of the XRT energy range.  Therefore, there should be 
a very small difference in the spectrum and light curves
between the $0.3-0.8$ keV and $0.8-10$ keV bands during 12.5 h
to about 21 h if our interpretation that there is some 
curvature in the spectrum is correct.   

\section{Early high energy data}

Using the parameter space we have obtained in 
Section 4.1 from late times radio, optical and X-ray data, 
we can calculate the external forward shock 
flux at high energies ({\it Fermi}/LAT band) at early times.
We choose to calculate the flux at 50 s and at 100 MeV.  If the 
LAT emission has an external forward shock origin, then 
this flux should be consistent with the 
observed value of 220 nJy (Abdo et al. 2009).  On the other hand, the GBM band flux 
($\sim 100$ keV) is dominated by the typical prompt variable 
source, whose origin remains uncertain; this emission lasts
for $\sim 30$ s and then it exhibits a sharp decay in the flux.  
The external forward shock emission at 100 keV and 
50 s should be smaller than the observed 100 keV 
flux, which is $\sim 0.4$ mJy (Abdo et al. 2009), so that the external forward shock flux does not 
prevent the observed flux to decay very rapidly ($\propto t^{-3}$)
as it is observed.     
 
\begin{figure}
\begin{center}
\includegraphics[width=12cm, angle = 0, clip=true, viewport=.0in .0in 8in 4.5in]{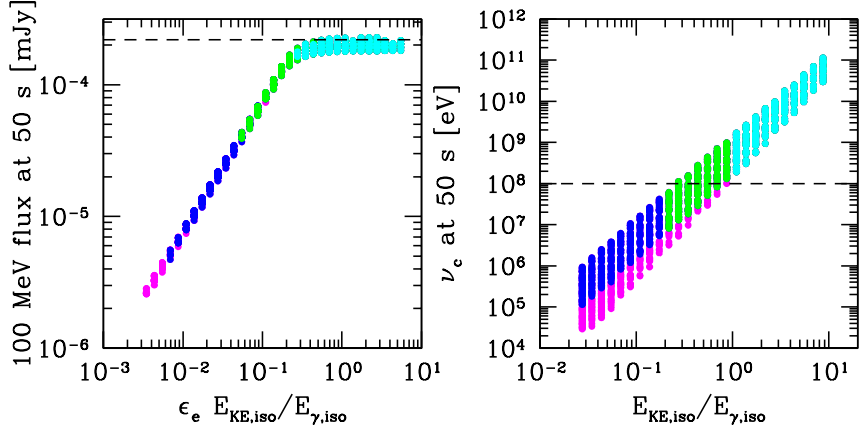}
\end{center}
\caption{100 MeV flux as a function of $\epsilon_e E_{KE,iso}/E_{\gamma,iso}$ 
and $\nu_c$ at 50 s as a function of $E_{KE,iso}/E_{\gamma,iso}$ in the left and 
right panels, respectively (for $p=2.2$).  The dashed lines indicate the 
observed flux at 50 s and $\nu_c=100$ MeV, respectively.  The injection 
frequency at 50 s is $\nu_i \sim 0.8$ keV (3 keV) for $p=2.2$ ($p=2.5$).  
The color coding is the same as in Figure \ref{fig4}.
(Please, see online version for color figure.)}
\label{fig5} 
\end{figure}

The calculation of the external forward shock emission at 
early times is not straightforward, since we find 
that the spectrum of injected electrons has a curvature. 
This should be taken into account when extrapolating 
from about a day to a few tens of seconds. Using 
the estimated evolution of $\nu_b$ as $\propto t^{-1.8}$
(Section 4.2.1), 
one finds that $\nu_b$ is $\sim 300$ MeV at 50 s. 
However, the temporal evolution of $\nu_b$ is highly uncertain and we do not 
know if it monotonically extends to very early times.  For this reason, 
we use two different values of $p$ when extrapolating from $\sim 1$ d to 50 s: 
$p=2.2$, consistent with the $p$ used before and 
$p=2.5$, which is the average of $p_{opt}$ and $p_x$ obtained 
in Section 4.2.  The results of these calculations are presented in 
Figure \ref{fig5}.

We find a large range of values for the external forward 
shock flux at 100 MeV and 50 s and plot this flux against 
$\epsilon_e E_{KE,iso}/E_{\gamma,iso}$; we choose this parameter 
because the flux above $\nu_c$ is roughly proportional
to $\epsilon_e E_{KE,iso}$ (see eq. \ref{eq:flux}) and we normalize 
$E_{KE,iso}$ to $E_{\gamma,iso}$.  Taking $\epsilon_e > 0.2$
as found for many GRB afterglows by Panaitescu \& Kumar (2001b) and 
$E_{KE,iso}/E_{\gamma,iso} > 1$ (so that $\eta < 0.5$), suggests that 
$\epsilon_e E_{KE,iso}/E_{\gamma,iso} > 0.2$, and in that case the 
calculated 100 MeV flux is consistent with the observed value to within
$\sim1$-$\sigma$ (for $p=2.2$).  For $p=2.5$ and $\epsilon_e E_{KE,iso}/E_{\gamma,iso} > 0.2$
the 100 MeV flux at 50 s is similar to the one obtained for $p=2.2$, 
but shifted downwards by only a constant factor of $\sim 2$.
This is a remarkable agreement given the 
fact that we have extrapolated the afterglow data from 
$\sim 1$ d to 50 s, and from radio, optical and X-ray to 100 MeV.

It is also important to know the location of the 
cooling frequency at 50 s, which would allow us to 
determine the spectrum and temporal decay index of the 
$> 100$ MeV light curve.  For both values of $p$, the cooling frequency 
is identical, and we plot it as a function of $E_{KE,iso}/E_{\gamma,iso}$
in Figure \ref{fig5}.  We can see that for 
$E_{KE,iso}/E_{\gamma,iso} > 1$, $100$ MeV$\lae \nu_c \lae$100 GeV 
at 50 s, therefore, one expects three possibilities: (i) $\nu_c \sim 100$ MeV
and it will very shortly fall below the 100 MeV band, 
and the $> 100$ MeV spectrum in this case should be consistent with being  
$\propto \nu^{-p/2}$, (ii) $\nu_c$
is much higher than 100 MeV and will thus remain above it
until $10^3$ s, which marks the end of the LAT observations and (iii) 
$\nu_c > 100$ MeV and it will cross the 100 MeV band during the observations. 
For (i), the 100 MeV light curve would 
be $f_{\nu} \propto t^{-1.15}\nu^{-1.10}$ for $p=2.2$
or $f_{\nu} \propto t^{-1.38}\nu^{-1.25}$ for $p=2.5$. 
For (ii) the 100 MeV light curve would 
be $f_{\nu} \propto t^{-0.90}\nu^{-0.60}$ for $p=2.2$
or $f_{\nu} \propto t^{-1.13}\nu^{-0.75}$ for $p=2.5$, 
and, lastly, for (iii) the light curve will transition 
from (i) to (ii).  

The observed $0.1-300$ GeV light curve for GRB 090902B 
decayed as $t^{-1.5 \pm 0.1}\nu^{-1.1\pm0.1}$ in 
the time interval $25-1000$ s (Abdo et al. 2009).  This light 
curve could be explained by scenario 
(i) above, for $p=2.5$, within $\sim1$-$\sigma$. 
This agrees nicely with the results shown in Figure \ref{fig4}; one 
can see that the 100 MeV flux increases almost linearly with 
$\epsilon_e E_{KE,iso}$ as expected for $\nu_c < 100$ MeV 
 and then reaches a plateau. The reason for this 
plateau is that it corresponds to $\nu_c \gae 100$ MeV 
at 50 s, and the flux below $\nu_c$ has already been precisely fixed by the 
X-ray flux at 12.5 h (one of the constraints), therefore, the 
100 MeV flux would also be fixed precisely. Shortly 
after 50 s, the $> 100$ MeV spectrum is consistent with being above 
$\nu_c$ as suggested by the data. Scenario (ii) above can be ruled out, 
because the LAT light curve decreases faster than predicted and also 
the predicted spectrum is too shallow compared with the observed one. 
Finally, scenario (iii) can also be ruled out, since a break in the LAT 
light curve was not detected in the data for GRB 090902B
(Abdo et al. 2009). It is worth mentioning that there is a small 
steepening of the $>100$ MeV light curve by, at most, $\sim t^{-0.03}$
when $\nu_c < 100$ MeV, due to the increase of $(1+Y)$ for electrons radiating at 100 MeV.

The flux from the external forward shock at 100 keV, 
which decays as $\sim t^{-1}$ would dominate the 
observed 100 keV light curve 
and prevent it from decaying quickly as it is observed 
($\sim t^{-3}$) unless the external shock contribution to the 100 keV flux 
is much smaller than the observed flux. To check this, we calculate the external forward 
shock flux at 100 keV and 50 s and find it to be $\sim 0.01$ mJy (for both 
$p$ values), which is a factor 
of $\sim 40$ smaller than the observed value.   This allows the observed 
prompt 100 keV flux to decay rapidly as observed. 
 
\section{The additional power-law component extending to 10 keV}

The GRB 090902B spectrum at early times displays a Band function 
in the sub-MeV energy range.  Time-resolved spectral 
analysis also shows a significant power-law component 
that appears to extend from the GeV range to the 
lowest energies ($\sim 10$ keV) and it is more intense than the Band 
function both for photon energy $\lae 50$ keV and $> 100$ MeV (Abdo et al. 2009). 
The Band function is usually associated with the
prompt emission and its origin remains uncertain.   
We address here the question whether the extra power-law 
component in the spectrum could have an external forward
shock origin.  We do this at 7 s,  which is the midpoint of interval {\it b} in 
the analysis of Abdo et al. (2009), where the power-law 
contribution is best constrained and its 
spectrum is $\beta = 0.94 \pm 0.02$.  The observed flux 
at 10 keV and 7 s is $\sim 12$ mJy.  

For the external shock origin of the power-law component at 7 s, the 
injection frequency should be below $\sim 10$ keV 
and the cooling frequency should be above the highest 
photon energy detected at that time, which is a few GeV.
Using the same procedure as in the last section we calculate 
$\nu_i$, $\nu_c$ and the flux at 10 keV at 7 s\footnote{Note that 
the $>100$ MeV light curve starts decaying at $\sim 10$ s,
which might correspond to the beginning of the deceleration 
phase of the external forward shock, however, 
extrapolating back to 7 s does not introduce a significant error.}.
We find that the injection frequency 
has a very narrow range of allowed values, $\nu_i \sim 15$ keV 
and $\nu_i \sim 60$ keV for $p=2.2$ and $p=2.5$, respectively.
The cooling frequencies at 7 s for $p=2.2$ and $p=2.5$ are identical
and are just a factor of $(50/7)^{1/2} \sim 2.7$ higher than 
the value at 50 s (see Figure \ref{fig5}), that is, 
$300$ MeV$\lae \nu_c \lae 300$ GeV at 7 s for $E_{KE,iso}/E_{\gamma,iso} > 1$.  
However, $\nu_c$ at 7 s cannot be larger than a few GeV, 
otherwise it will stay above 100 MeV for the entire 
duration of the LAT emission, which is inconsistent with the 
observed spectrum during this period.  
In addition, the expected spectrum between 
$\nu_i$ and $\nu_c$ would be $\beta = 0.60$ and $\beta = 0.75$
for $p=2.2$ and $p=2.5$, respectively, which is inconsistent with 
the observed value.  Finally, the expected flux at 10 keV and 7 s
lies in a very narrow range, and it is $\sim 0.3$ mJy (for both $p$ values), 
which is a factor of $\sim 40$ smaller than the observed value.
All these arguments suggest that the power-law detected in addition 
to the Band spectrum at early times is unlikely produced by the 
external shock, at least in the simplest version of this 
model.


\section{Discussion}

In this paper we have considered the late afterglow data 
of GRB 090902B in the context of the external forward shock model.
The optical data is entirely consistent with this model.  However, 
the X-ray flux decays faster than expected for the observed 
X-ray spectrum.  We consider three possibilities that could have 
steepened the X-ray light curve falloff.

First, we considered radiative losses or evolving microphysical 
parameters that might steepen the X-ray flux.  However, 
since the optical and X-ray fluxes have very similar dependence 
on these parameters, it is not possible to steepen the X-ray light 
curve and at the same time not steepen the optical light curve, that is, 
obtain a solution consistent with the observations. 
Second, we carried out a very
detailed calculation of the Compton-$Y$ parameter in hopes 
that its increase with time could steepen the X-ray data
if the cooling frequency is below the X-ray band. However, we find that its 
effect on the steepening of the light curve is extremely small and 
unable to explain the observed data.

Another modification of the standard external forward shock 
model we considered is the possibility that the spectrum of injected 
electrons at the shock front is not a single power-law, 
but that it exhibits some curvature.  We find that this 
is the only modification that can reconcile the 
theoretical expectations with the afterglow data of 
GRB 090902B.  

The curvature we find in the spectrum of injected electrons might 
not be inconsistent with the recent particle-in-cell simulations of particle acceleration 
in relativistic collisionless shocks (Sironi \& Spitkovsky 2011). 
Simulations need to be able to run for longer to achieve time-scales and 
energy range relevant to the GRB afterglow.  Future simulations should be 
able to explore our claim of the presence of curvature in the electron 
energy distribution.  The slight curvature, invoked phenomenologically, 
is tied to the physics of particle acceleration, which suggests in this 
case that the higher energy electrons -- those radiating X-rays -- are 
accelerated slightly less efficiently than those radiating optical photons.  
The fact that we find a ``downward'' curvature ($p_x > p_{opt}$) is reassuring, 
since the opposite would have been contrived. 

The synchrotron frequency that corresponds
to the electrons Lorentz factor at which the spectrum 
curves should be at $\sim 0.8$ keV at the beginning 
of the {\it Swift}/XRT observations.  
The break frequency decreases with time as $\sim t^{-1.75}$
and it falls below the XRT band at about 21 h (Section 4.2.1). 
An inspection of the 0.3-1.5 keV XRT light curve shows a 
slight flattening during this time in this light curve compared to 
the 1.5-10 keV one, just as predicted (Evans et al. 2007).
A careful analysis of the XRT data shows that the 0.3-1.5 keV
data are well fit by a broken power-law as predicted, but the presence
of this break in the light curve cannot be claimed with high 
statistical significance; the  
predictions of this model however cannot be ruled out due to the sparseness 
of the data and the fact that the theoretical decay slopes before 
and after the break ($t^{-0.90}$ and $t^{-1.35}$) are similar
(Margutti, personal communication).  Similarly, the 
possible break in the spectrum ($\nu^{-0.6}$ to $\nu^{-0.90}$) 
could not be found since the errors 
in the spectrum are on the order of the difference between 
the two different indices (before and after the break) that 
one expects. Moreover, the theoretically expected change in the spectrum and 
decay indices appears at the lower energy band of XRT, which is 
affected by absorption.  There is the possibility, 
however, to use the absorption calculated using the optical band 
to fix the level of absorption in the X-ray band and look for this 
small change in the spectrum and temporal decay indices.

This work assumes that the late afterglow in all bands is produced 
by the same population of electrons.  Can we abandon this 
scenario and invoke one in which, for example, 
the X-ray and optical data are produced by two different 
sources? It seems unlikely.  We discuss this below.

Let us start with the assumption that the X-ray flux is originated in 
the external forward shock, that is, by the interaction 
of the GRB jet with the CSM.  The optical data might 
have another origin. However, since the optical flux decays as a 
single power-law for a long period of time, its origin 
is also some form of an external shock. This external shock does not 
have to necessarily be the same one that produced the 
X-ray afterglow. It 
could be, for instance, that a lower LF cocoon material
interacts with the CSM and drives an external shock
which produced the optical radiation.
As can be seen in Section 4.2.1, the optical and 
the X-ray fluxes close to 1 d fall on 
a single power-law spectrum which is roughly 
consistent with the observed optical and X-ray spectra.
If we assume that for these two shocks $\epsilon_e$,
$\epsilon_B$, and $n$ are approximately equal, then $E_{KE,iso}$ 
should also be the same.  Why should $E_{KE,iso}$ for the two
different and unrelated sources be the same?  
For this reason, we think that this possibility is  
contrived and suggest the same origin for both X-ray and optical 
photons.  

Even if X-ray and optical photons are produced by the 
same population of electrons: Could it be that optical (and radio) 
photons are produced via synchrotron while X-ray photons 
originate via Inverse Compton radiation scattering 
of synchrotron photons?  This interesting 
possibility can be ruled out.  At $\sim 1$ d, if we extrapolate 
the optical flux using optical spectrum to radio band we 
overestimate the observed radio flux by a factor of $\sim 700$.
This means that there is a break in the spectrum, which
would correspond to $\nu_i$, at $\sim 10^{-2}$ eV.  
At this frequency, the peak synchrotron flux is $\sim 1$ mJy.
The optical depth to Thompson scattering in the external 
forward shock at $\sim 1$ d is $\tau_e \sim \sigma_T n R \sim 10^{-7}$, 
while the ratio of X-ray flux to the peak synchrotron flux 
at the same time is $\sim 10^{-4}$.  Therefore, this 
possibility can be safely ruled out since the X-ray
flux is at least $10^3$ times larger than the maximum expected
flux for the synchrotron-self-Inverse Compton process. 

We have also found that when we constrain the 
X-ray, optical and radio fluxes at $\sim 1$ d one finds 
a large allowed range for $\epsilon_B$.   
When we further constrain $\epsilon_e > 0.2$, as found 
by Panaitescu \& Kumar (2001b) and take the 
isotropic kinetic energy in the blast wave to be larger 
than the isotropic radiated gamma-ray energy during the 
prompt phase ($E_{KE,iso} > E_{\gamma,iso}$), then the 
magnetic field in the source is consistent with being produced 
via shock-compressed CSM field as shown in Figure \ref{fig4} and 
suggested by Kumar \& Barniol Duran (2009, 2010), Barniol Duran \& Kumar (2010).  
The required seed field -- the upstream field -- 
before compression is $\lae 10 \mu$G. The constraint 
$E_{KE,iso} > E_{\gamma,iso}$ is applied in order to avoid a radiative 
efficiency larger than 50\%; however, even
when we take $E_{KE,iso} > E_{\gamma,iso}/5$ 
-- corresponding to a radiative efficiency of 
80\% -- we still find that the shock compression 
scenario holds with just a slightly larger seed field of $\lae 60\mu$G
(see Figure \ref{fig4}).   

To obtain the region of $\epsilon_B$-$n$ plane allowed by the GRB 090902B
data we have used the X-ray, optical and radio fluxes at $\gae 1$ d, 
since the external reverse shock might dominate the optical data 
until $\sim 0.7$ d (see fig. 1 of Pandey et al. 2010).  
There is the possibility, however, that the radio flux 
at $\sim 1$ d might still be dominated by the external reverse
shock, since the external reverse shock at the radio band 
might decay slower than at the optical band. This is the reason why 
we have obtained the $\epsilon_B$-$n$ plane using the radio flux 
at 5.6 d (Cenko et al. 2011).  We expect that at this time 
the contribution from the reverse shock to the radio flux 
is negligible.  Nevertheless, if instead of using the radio 
flux at 5.6 d we use the radio flux measured 
at 1.3 d at the Westerbork Synthesis Radio Telescope (van der Horst et al. 2009),
we find that our general results are not modified: the requirements
for the magnetic field are the same; however, the density 
increases by a factor of $\sim 10$; points shown in Figure \ref{fig4} 
are shifted to the right by a factor of $\sim 10$ along the 
diagonal bands in this case.  

The discrepancy in the value of $\epsilon_B$ between 
our work and the work of other authors, e.g. Liu \& Wang (2011), arises from 
the fact that these authors assume that the cooling frequency is 
between the optical and X-ray frequencies.  We make no such 
a priori assumption (the only assumption we make is that 
X-ray, optical and radio data come from synchrotron emission in the 
external forward shock) and, in fact, our very detailed analysis shows 
that if the cooling frequency is between the optical and X-ray bands, 
it is impossible to have the X-ray temporal decay index match the observations.  
Liu \& Wang (2011) find a value of $\epsilon_B \sim 10^{-3}$ and a density of 
$n \sim 3\times10^{-4}$ cm$^{-3}$ (for the wide component of their two-component jet, 
which dominates the flux at late times), which may give the impression that their $\epsilon_B$ 
is $\sim 10^3$ larger than our found value.  However, this is not so. For the 
$\epsilon_B$-n region where $\epsilon_e > 0.2$ and $E_{iso}/E_{\gamma,iso} > 1$, 
the maximum allowed value of $\epsilon_B$ we find for this value of $n$ is 
$\epsilon_B \sim 4\times10^{-5}$ (see Fig. \ref{fig4}), which is 30 times smaller than the value 
reported by Liu \& Wang (2011).  The discrepancy is still large, but not at 
the level of a factor of $10^3$ and -- as discussed -- originates 
from the incorrect assumption of the location of the cooling frequency. Note 
that Cenko et al. (2011) do not find $\epsilon_B$ for GRB 090902B, 
but instead assume equipartition and fix their value of $\epsilon_B$.

\subsection{Electron acceleration and the upstream magnetic field}

\begin{figure}
\begin{center}
\includegraphics[width=12cm, angle = 0, clip=true, viewport=.0in .0in 8in 4.5in]{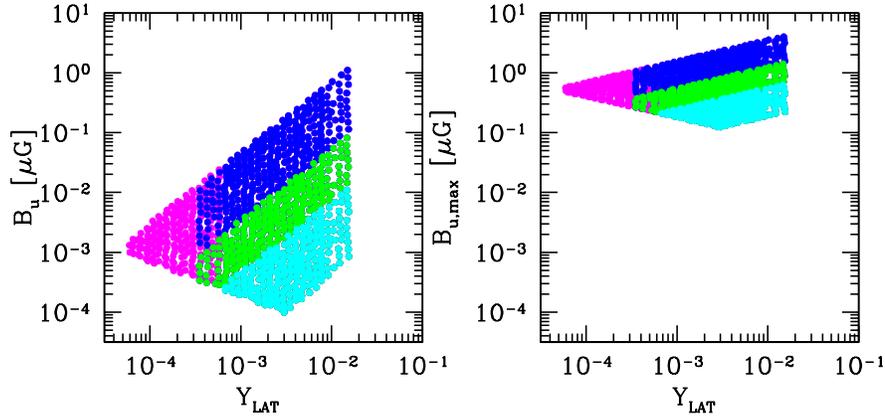}
\end{center}
\caption{Lower limit on the upstream circumstellar medium magnetic field, $B_u$, as a 
function of the Compton-$Y$ parameter for electrons 
radiating at 100 MeV at $10^3$ s, $Y_{LAT}$ (left panel).  $Y_{LAT}$ is 
calculated using the method described in Section 3 for the 
allowed subspace of the 4-D parameter space ($\epsilon_e$, $\epsilon_B$, $n$, $E_{KE,iso}$) for 
GRB 090902B afterglow data shown in Figure \ref{fig4} (for $p=2.2$).  $B_u$ was 
obtained by requiring that the Inverse Compton cooling
time for these electrons is larger than their acceleration timescale: Method 1.
We also calculate a lower limit on $B_u$ by requiring that the 
electrons radiating at 100 MeV at $10^3$ s are confined to the shock front: Method 2
($B_{u}$ obtained by this method does not depend on $Y_{LAT}$ and has an extremely weak
time dependence). The maximum of these two values, $B_{u,max}$, is the true minimum of 
$B_u$, which for GRB 090902B is given by Method 2 (right panel) and it is plotted 
as a function of $Y_{LAT}$ to aid in the comparison between the two panels; $B_{u,max} \lae 5$ $\mu$G.
The color coding is the same as in Figure \ref{fig4}.
(Please, see online version for color figure.)}
\label{fig6} 
\end{figure}

Recently, the detection of 100 MeV photons up to $\sim 10^3$ s
in {\it Fermi} GRBs has been used to calculate a lower limit 
on the upstream CSM field by requiring that the Inverse Compton 
loss timescale is larger than the acceleration timescale
of electrons radiating at 100 MeV (Li \& Zhao 2011). For the subspace of 4-D 
parameter space ($\epsilon_e$, $\epsilon_B$, $n$, $E_{KE,iso}$) consistent 
with the data for GRB 090902B (Section 4.1) we calculate  
Compton-$Y$ for electrons radiating at 100 MeV at $10^3$ s
(using $p=2.2$), and use that to determine 
a lower limit to the upstream field (Figure \ref{fig6}).
We calculate another lower limit to the 
upstream field by requiring that electrons radiating in the 
LAT band are confined to the shock front (Piran \& Nakar 2010, 
Barniol Duran \& Kumar 2011); we determine this lower limit for electrons 
radiating at 100 MeV and $10^3$ s (see equation (3) in Barniol Duran \& Kumar 2011).
The true lower limit to the upstream magnetic field is taken to be the 
larger of these two limits and the result is shown in Figure \ref{fig6}.  
We find that for GRB 090902B the electron confinement requirement 
gives a larger upstream field for much of the allowed parameter space, 
and an upstream field of $\sim 5 \mu$G is sufficient to confine electrons
producing 100 MeV radiation and to avoid excessive Inverse Compton 
losses while electrons are traveling upstream of the shock front.

The lower limit on the upstream CSM magnetic field we find is 
much smaller than the one found by Li \& Zhao (2011).  This discrepancy 
arises because of two points: 1. The difference between the  
calculation of Inverse Compton loss --  Li \& Zhao (2011)
used a simplified calculation for Inverse Compton loss
whereas we have carried out an almost exact numerical calculation  
as described in Section 3.2.1, and 2. Different numerical factors 
in their equation of the lower limit on the upstream field\footnote{The equation of Compton-$Y$ described 
in the Appendix of Li \& Zhao (2011), eq. (B6), overestimates 
$Y_{LAT}$ by a factor of $\sim 8$ by: (i) ignoring the integral over incoming photon angle 
and (ii) by using the total Klein-Nishina cross-section, whereas the
differential cross-section is needed, since the energy transfer 
to the photon depends on the outgoing photon angle. (iii) Also, they 
use a pre-factor of $0.3$ in our equation (\ref{eq:nu2}), inconsistent
with Wijers \& Galama (1999), that also slightly overestimates 
$Y_{LAT}$ by a factor of $\sim 2$.  (iv) Their expression of the acceleration time, $t_a$, is 
larger than ours by a factor of $5$.  (iii) and (iv), combined, 
overestimate the lower limit on $B_{u}$ in Li \& Zhao (2011) eq. (28) by increasing the 
numerical factor of this equation by $5/0.3\sim17$.  Nevertheless, even if we increase $B_{u}$ 
by a factor of $5$ in the left panel of Fig. \ref{fig6}, to agree with their $t_a$, the right panel 
of Fig. \ref{fig6} would remain unchanged.}. 


We note that the $\epsilon_B$-$n$ parameter 
space shown in Figure \ref{fig4} should be slightly revised 
to reflect our findings in Figure \ref{fig6}; results in 
Figure \ref{fig4} did not include 
the constraints on magnetic field determined in this subsection. 
However, the modification to the lower limit of the CSM field in the 
$\epsilon_B$-$n$ parameter space is very small, and it is non-existent
for the case when $E_{KE,iso}/E_{\gamma,iso}>1$ and $\epsilon_e > 0.2$, 
therefore, we have left Figure \ref{fig4} unchanged.


\section{Conclusions}

We have analyzed the late time afterglow X-ray, optical and 
radio data for GRB 090902B in the context of the 
synchrotron radiation mechanism in the external forward shock
model.  We find that a curvature in the power-law electron energy 
distribution is needed in order to provide a good fit to the late time 
optical and X-ray data; radiation losses, varying microphysical 
parameters and an increase in Inverse Compton losses all fail 
to explain the observed data.  The late time afterglow fit gives an $\epsilon_B$ (fraction of total 
energy in the shock imparted to magnetic fields) consistent 
with shock-compressed circumstellar medium magnetic field 
of $\lae 10$ $\mu$G and $\lae 60$ $\mu$G if we take the 
efficiency for producing gamma-rays to be $\sim 50\%$ and 
$\sim 80\%$, respectively.  

Particle acceleration in the external forward shock 
allows us to set a lower limit on the upstream 
circumstellar medium magnetic field.
We find that the field strength 
in the unshocked medium in the vicinity of GRB 090902B must be at least 
2 $\mu$G in order to produce 100 MeV photons at 
50 s$\lae t \lae 10^3$ s (Barniol Duran \& Kumar 2011).  

The calculation presented in this paper represents 
an improvement in our previous afterglow modeling 
due to a more precise calculation of Inverse Compton 
losses.  Here, we include Klein-Nishina effects and also 
relativistic corrections of the outgoing energy of the Inverse 
Compton scattered photons.  We also calculate the electron 
energy distribution self-consistently by determining 
the synchrotron emission and using it to determine the 
Inverse Compton losses, which in turn modify the electron 
energy distribution. 

The flux calculated at 100 MeV at 50 s using the 
external forward shock parameters obtained from 
the late afterglow data is consistent with the 
{\it Fermi}/LAT data, confirming our previous 
claims (Kumar \& Barniol Duran 2009, 2010).
We also calculated the expected external forward shock 
at 100 keV and 50 s.  At this time, the observed 100 keV
light curve is undergoing a fast decay ($\sim t^{-3}$).  We find that the external 
forward shock at 100 keV and 50 s is smaller than the 
observed value by a factor of $\sim 40$, easily allowing 
the observed light curve to decay quickly (confirming 
our earlier results in Kumar \& Barniol Duran 2010).  We speculate  
that for some small fraction of GRBs this steep decay will not be seen, 
instead, one will see a smooth slowly decaying ($\sim t^{-1}$)
light curve emerge after the main prompt, variable, emission
is over.

The origin of the high-energy emission at $<50$ s
(when the prompt GBM phase is active) is still a subject of debate.
It is claimed that GRB 090902B exhibits variability in the LAT band
(Abdo et al. 2009), which would make the external forward shock 
origin for $<50$ s difficult, since this model in its simplest version 
predicts smooth light curves. However, the fact that for this GRB the 
$>50$ s 100 MeV light curve connects smoothly -- as a single power-law --
with the $<50$ s one does argue in favor of the external forward shock 
origin (see, however, Maxham, Zhang \& Zhang 2010).  
The delay on the detection of the first 100 MeV photons with respect of the 
first GBM photons from this GRB can also be explained by the external forward 
shock model (Barniol Duran \& Kumar 2011). A dedicated study of the variability 
during the prompt phase in the LAT band would certainly shed more 
light on whether the prompt ($<50$ s) 100 MeV emission of this GRB 
can be originated from the external forward shock.
 

We find that the cooling frequency at $\sim 50$ s is $\sim 100$ MeV
for GRB 090902B.  If this is correct, one should be able to track 
the cooling frequency as it passes through the LAT band at earlier times 
($\nu_c \propto t^{-1/2}$).  A presence of this behavior in the data 
of this and other LAT detected GRBs would help confirm an external shock 
origin for the LAT emission.

Finally, the prompt spectrum of GRB 090902B contains a power-law component in addition 
to the usual Band function.  We argue against the external shock origin of the extra power-law 
component at 7 s (Section 6).  

\section*{Acknowledgments}
RBD dedicates this work to Don Angelo Botta, and thanks 
Jessa Barniol for her support during the writing of this manuscript.
RBD thanks useful conversations with Ramesh Narayan, Eliot Quataert, Dale Frail, 
Brad Cenko, Raffaella Margutti, Cristiano Guidorzi, Patrick Crumley, Rodolfo Santana 
and Chris Lindner. We thank the referee for a constructive report.
RBD gratefully acknowledges support from the Lawrence C. Biedenharn Endowment for
Excellence.  This work has been funded in part by NSF grant ast-0909110.
This work made use of data supplied by the UK Swift Science Data Centre at the University of Leicester.



\end{document}